\documentclass[usenatbib]{mn2e}

\usepackage{graphics}
\usepackage{epsfig}
\usepackage{natbib}

\begin{document}

\title[COLD GASS: Comparison with semi-analytic models]{COLD GASS, an IRAM Legacy Survey of Molecular Gas in Massive Galaxies:  III.  Comparison with semi-analytic models of galaxy formation }

\author[G. Kauffmann et al.]{Guinevere Kauffmann$^{1}$\thanks{E-mail: gamk@mpa-garching.mpg.de}, Cheng Li$^{2}$,
Jian Fu$^{1,2}$, Am\'{e}lie Saintonge$^{3}$,  Barbara Catinella$^{1}$,     
\newauthor
Linda J. Tacconi$^{3}$,  Carsten Kramer$^{4}$, Reinhard Genzel$^{3}$,
Sean Moran$^{5}$, David Schiminovich$^{6}$\\   
$^{1}$Max-Planck Institut f\"{u}r Astrophysik, 85741 Garching, Germany\\
$^{2}$Max-Planck-Institut Partner Group, Shanghai Astronomical Observatory\\
$^{3}$Max-Planck Institut f\"{u}r extraterrestrische Physik, 85741 Garching, Germany\\
$^{4}$Instituto Radioastronom\'{i}a Milim\'{e}trica, Av. Divina Pastora 7, Nucleo Central, 18012 Granada, Spain\\
$^{5}$Johns Hopkins University, Baltimore, Maryland 21218, USA\\
$^{6}$Department of Astronomy, Columbia University, New York, NY 10027, USA}

\maketitle

\begin{abstract} 
We compare the semi-analytic models of galaxy formation of Fu et al. (2010),
which track the evolution of the radial profiles of 
atomic and molecular gas  in galaxies, with
gas fraction scaling relations derived from a
stellar mass-limited sample of 299 galaxies from the COLD GASS
survey. These galaxies have measurements of the  
CO(1-0) line from the IRAM 30-m telescope
and the HI line from Arecibo, as well as measurements of star formation
rates, stellar masses, galaxy  sizes and concentration parameters 
from GALEX+SDSS photometry.
The models provides a good description of
how condensed  baryons in star-forming galaxies 
are partitioned into atomic and  molecular
gas and stars as a function of galaxy stellar mass 
and stellar surface density.  The models   
do not reproduce the observed tight relation between
stellar surface mass density  and bulge-to-disk ratio in these galaxies. 
The current implementation of ``radio-mode feedback'' in the models produces     
trends that disagree strongly with the data. 
In the models, gas cooling shuts down in nearly all galaxies in  dark matter halos
above a mass of $\sim 10^{12} M_{\odot}$. As a result,  
stellar mass is the   observable that best predicts whether
a galaxy has little or no neutral gas, i.e. whether a galaxy has been quenched.
In contrast, our data show that
quenching  is largely {\em independent} of  stellar mass. 
Instead, there are clear thresholds in 
bulge-to-disk ratio and in stellar surface density that demarcate the location
of quenched galaxies in our chosen parameter space.
We speculate that the disagreement between 
the models and the observations may be resolved if radial transport of gas      
from the outer disk is included as an additional 
bulge-formation mechanism in the models. In addition, we propose that
processes associated with
bulge formation  play a key role in depleting the neutral gas in galaxies and 
that gas accretion is  suppressed in a significant fraction of
galaxies following the formation of the bulge, even in dark matter halos of low mass.
\end{abstract}


\section{Introduction}

An important goal in modern galaxy formation is an improved
understanding of the physical processes  
that regulate the rate at which stars form in galaxies. 
These processes include the cooling and accretion of gas within
dark matter halos, the transformation of the accreted gas into 
molecular clouds and stars, and the effects of ``feedback'' from 
massive stars and accreting black holes on the gas in and around galaxies.

The Lambda-Cold Dark Matter model provides detailed predictions for
how  dark matter halos  
assemble over time. Within these halos, gas will cool and settle into
a rotationally-supported disk. One hotly debated issue is the degree
to which gas loses angular momentum to the surrounding dark matter
during this process. In many ``semi-analytic'' models, the angular 
momentum of the gas is  assumed to be conserved. This
produces a  size-circular velocity
relation for galactic disks that is in good agreement with observations (Mo ,Mao \& White 1998).
More detailed numerical hydrodynamical simulations of disk galaxy
formation in a $\Lambda$CDM universe demonstrate, however, 
that the amount of angular
momentum that is lost during the collapse of the dark matter 
halo depends sensitively on how  feedback processes are  included
in the models (Navarro \& Steinmetz 1997; Governato et al 2004,2007). 
In many simulations, the disks form early and  are 
too compact.  The resulting star formation timescales
are then too short to be consistent with observations (e.g. Oser et al 2010). 

Another important unsolved problem is to understand why galaxies divide 
into two distinct ``families'' -- those with ongoing active star formation,
and those where star formation has been quenched (Strateva et al 2001;
Kauffmann et al 2003b;  Baldry et al 2004). It has now become 
clear that these distinctions existed already
at redshifts as high as $2.5$ (Williams et al 2010; Wuyts et al 2011),
when cold gas accretion rates in almost 
all dark matter halos are predicted to be very high. 

Many  recent theoretical galaxy formation models that aim to
reproduce the statistical properties of the massive galaxy population 
(e.g. Croton et al 2006; Bower et al 2006; Cattaneo et al 2006 De Lucia \& Blaizot
2007; Somerville et a 2008; 
Guo et al 2011; Lu et al 2011) invoke star-formation quenching mechanisms
that set in at a characteristic dark matter halo {\em mass}. The 
characteristic mass is associated with the transition between the 
regime where gas cooling times are short compared to the dynamical
time of the dark matter halo and gas accretes in the form  cold,
condensed clouds,
and the regime where cooling times are long and gas accretes from a
corona of gas that is in virial equilibrium with the surrounding halo. 

One might ask {\em why} lowered rates of star 
formation in galaxies should be linked with this transition.
One argument that is often used 
is that relativistic jets, 
generated  when gas accretes onto black holes, heat
the surrounding hot gas and prevent it from forming stars 
(see, for example, Croton et al. 2006).    
The conditions under which black holes produce jets are not well understood.
There is clear observational evidence that radio-emitting jets play a role
in regulating the cooling of gas in nearby galaxy clusters (see McNamara \& Nulsen 2007 for a recent review). 
Theoreticians have thus made  an ``ansatz''  that jets from  radio-loud AGN will
prevent gas from cooling and forming stars in {\em all} dark matter 
halos that are predicted to contain a hot gas atmosphere.
The characteristic halo mass threshold that separates  
star-forming and quiescent galaxies is the main
factor that determines how massive galaxies evolve in current models,
so it is clearly important to test the existence of such a threshold using
real observations.

We note that accretion and quenching processes affect the  
{\em gas components} of galaxies. The most direct empirical constraints on  
how they operate thus come from observations of the gas 
in galaxies. This has been the primary motivation for      
the Galex Arecibo SDSS  survey (GASS), as well as the CO Legacy Database 
for GASS (COLD GASS), which are measuring the atomic and molecular
gas contents of an unbiased sample of several hundred galaxies with
redshifts between 0.025 and 0.05, and with stellar masses 
in the range $10^{10}<M_*<10^{11.5} M_{\odot}$. In both surveys, 
the strategy is to observe each galaxy until the HI/CO lines are detected,
or until upper limits in the ratio of  atomic and molecular gas mass to stellar mass 
of $\sim 0.015 $ reached. The aim of the program is to carry
out a  census of the condensed baryons in galaxies in the local Universe,
to study scaling relations between the gas and stellar properties of
these galaxies, and to understand gas accretion
and quenching in galaxies. 

Details of the GASS and COLD GASS survey designs, as well as
target selection and observing procedures are given in Catinella et al (2010)
and Saintonge et al (2011a). These two papers also presented relations between
the HI and H$_2$ mass fractions of galaxies (defined as $M_{\rm HI}$/M$_*$
and $M_{\rm H_2}$/M$_*$) and global galaxy parameters such as stellar mass M$_*$,
stellar mass surface density $\mu_*$, galaxy bulge-to-disc ratio (as parameterized
by the concentration index $C$ of the $r$-band light), 
and specific star formation
rate SFR/M$_*$ (see also Schiminovich et al 2010). 
The two surveys uncovered  sharp thresholds in $\mu_*$ and $C$ 
below which all galaxies have a measurable cold gas component, 
but above which the detection rate of the CO and HI lines drops suddenly, 
suggesting that ``quenching processes'' have occurred in these systems.

In this paper, we compare the observed gas fraction relations
from the COLD GASS survey with predictions
from the semi-analytic models of Fu et al 
(2010; hereafter F10), which track the formation
of molecular gas in disk galaxies. 
The F10 models are based on 
the L-galaxies semi-analytic code, described in detail in Croton et al (2006) and
updated in De Lucia \& Blaizot (2007).

The models are currently
being configured to operate on the latest update of the L-galaxies code
by Guo et al. (2011);  the comparison in this paper will be
restricted to the version of the models in the published F10 paper. 
The main new aspect of these
models  is that galactic discs are represented
by a series of concentric rings in order to track the evolution in the
gas and stellar surface density profiles of galaxies over cosmic time.   
Two simple prescriptions for molecular gas formation 
processes are included: one is based on the analytic 
calculations by Krumholz, McKee \& Tumlinson (2009; hereafter KMT), 
and the other is a prescription where the H2 fraction is 
determined by the pressure of the interstellar medium (Blitz \& Rosolowsky 2006, hereafter BR), 
with an implementation similar to that in Obreschkow et al. (2009). 
The free parameters of the models
that regulate the rate at which gas is turned into stars and the  efficiency with
which supernovae reheat gas as a function of halo mass have been  tuned
to reproduce a number of key observables, including the observed
present-day galaxy  luminosity function, the gas-phase mass-metallicity relation, and
the mean HI gas fraction as a function of  B-band luminosity  
(see De Lucia, Kauffmann \& White (2004) for a more detailed discussion).

There have been a number of other cosmological models that follow the formation of molecular gas
in disks.
 Lagos et al (2011) have 
developed semi-analytic models that track molecular gas formation
in cosmological simulations of galaxy formation, but these models
do not predict the detailed radial profiles
of galaxies as in F10. Robertson \& Kravtsov (2008) and 
Dutton et al (2010) model the radial
profiles of the gas and stars in galaxies, but their models are not embedded
within full cosmological N-body simulations. Gnedin et al (2009)
have carried out very high resolution cosmological 
simulations of molecular gas formation
that include radiative transfer and detailed treatment of the
chemistry, but their box
sizes are too small to make statistical predictions. 

The F10 paper demonstrated that their models could fit the radial
HI, H$_2$, stellar mass and SFR profiles of spiral galaxies from the
THINGS/HERACLES surveys. In this paper, we investigate whether the models 
also yield correct gas fraction scaling relations and  distribution functions. As will be seen, it is
useful to break this analysis  into two distinct parts:
1)comparison between models and observations  
of the distribution of gas fractions in the  
population of galaxies with detectable gas, 
2) comparison of how {\em absence of gas}, which we refer to
as ``quenching'', is manifested
as a function of observable quantities such as stellar mass,
stellar surface density, and bulge-to-disk ratio.  As we will show,
such a comparison elucidates  those aspects of the
models which work well, and those that fail. Throughout this paper, we have
assumed a cosmology with $\Omega=0.3$, $\Lambda=0.7$  and $H_0 = 70$ 
km s$^{-1}$ Mpc$^{-1}$ to derive our observational quantities.
The cosmology assumed for the F10 model is a $\Lambda$CDM
models with  $\Omega=0.25$, $\Lambda=0.75$,   $H_0 = 73$
km s$^{-1}$ Mpc$^{-1}$ and $\sigma_8=0.9$.  

\section {Generating comparison samples from the simulations}

The semi-analytic models used in this paper are nearly
identical to those  described in detail in F10. We have made only  
one small change.
The free parameters in the  F10 models were chosen so that reasonable fits
to a variety of observables could be 
obtained. These  observables included the H$_2$ mass function of
Keres et al (2003) derived using galaxies drawn from the 
FCRAO Extragalactic CO Survey (Young et al 1995). These authors adopted a 
CO-to-H$_2$ conversion
factor that was a factor of 1.5 times larger than the one 
adopted by Saintonge et al (2011a).  
\footnote{$\alpha_{\rm CO} = 3.2 M_{\odot}$ (K km s$^{-1}$ pc$^2$)$^{-1}$, 
which does not
include a correction for the presence of helium.} 
To bring the simulations back into agreement with the H$_2$ data 
using the conversion factor adopted by Saintonge et al, we simply increase
the supernova reheating rate (given by the parameter $\epsilon_{disc}$) in Table
1 of F10 by a  similar factor ($\epsilon_{disc}=5$ instead of 3.5). 
We note that changing $\epsilon_{disc}$ simply shifts 
the amplitude of the gas mass fraction
scaling relations up or down. It does not change 
the predicted  slope or scatter in these
relations. We also note that the  $H_2$ mass function derived from the COLD GASS data
for galaxies with $\log M_* > 10^{10} M_{\odot}$ agrees well with that of 
Keres et al (2003) at the high mass end, once the same  CO-to-H$_2$ conversion
factor is adopted (Fu et al 2012, in preparation).         

The observational sample is an update to that described in Saintonge et al 2011(a),
consisting of 299 galaxies with CO(1-0) line observations with the IRAM 30 m telescope.
HI line measurements from the GASS survey (Catinella et al 2010) are available for
270 out of the 299 galaxies.
Because the observational sample is selected only by stellar mass, 
it is easy to compare simulation results directly with the
data. There are only two observational selection issues that require careful treatment:

\begin {enumerate}                       
\item In both the GASS and COLD GASS surveys, targets are selected so that the
resulting stellar mass distribution is approximately flat.
This was done in order to be able to compute gas fractions in different stellar
mass bins with roughly the same errors.
The sampling must be taken into account  when  we look at gas fraction
distributions as a function of parameters other than stellar mass.
In this paper, we adopt two approaches: a) We either weight each galaxy in
the  sample by the inverse of $f(M_*)$, where 
$f(M_*)$ is the  ``sampling rate function'' required to transform the
true stellar mass distribution to a flat one, b) We apply the sampling
rate function $f(M_*)$ to the simulated galaxies to create ``mock catalogue'' with
flat stellar mass distributions that can be compared directly with the data.    
\item It is important to take account of the detection limits of the survey when
comparing model galaxies with the data. This is illustrated 
in the top panels of Figure 1, 
where we plot M(HI)/M$_*$ (left) and M(H$_2$)/M$_*$ (right) as a function 
of stellar mass for 299 galaxies that have been observed in CO
as of July 2011 (270 of these have HI observations from the GASS survey). 
Galaxies where the HI or CO line was detected (see Catinella et al 2010 and
Saintonge et al 2011a for details about line
detection procedures) are plotted in blue. Galaxies without detections
are plotted in red at the positions 
of their 5$\sigma$ upper limits in M(HI)/$M_*$ or
M(H$_2$)/$M_*$.   

To a very close approximation, the HI mass fraction 
limit log [M(HI)/M$_*$]$_{lim}$ is -1.82 (corresponding to a HI mass
fraction limit 0.015 ) for galaxies with $\log M_* > 10.3$. For galaxies
with $10 < \log M_* < 10.3$, log [M(HI)/M$_*$]$_{lim} = -1.066 \log M_*+9.16$.

The CO line detection limits are somewhat more complicated. 
As discussed in Saintonge et al (2011a), the integration times are set so that 
 log [M(H$_2$)/M$_*$]$_{lim}$=-1.82 for galaxies with  $\log M_* > 10.6$. For lower mass galaxies,
integration times are nominally  set to reach a fixed r.m.s.   of around 1.1 mK per 20 km/s wide channel, but
the actual value fluctuated according to weather conditions at the time of observation.
Therefore, when creating our mock catalogues from the simulations, we simply impose a distribution of upper limits similar
to that seen in the top right panel of Figure 1. We take 
[M(H$_2$)/M$_*$]$_{lim,max}$, the {\em maximum} possible value of the  H$_2$ mass fraction limit, 
as    log [M(H$_2$)/M$_*$]$_{lim,max}$ = -1.72
if $\log M_* > 10.6$, and log [M(H$_2$)/M$_*$]$_{lim,max}$ = 8.78- log M$_*$  
if $\log M_* < 10.6$. 
The H$_2$ mass fraction limit that we adopt is randomly  distributed between the  maximum
value and a minimum value that is  0.35 dex smaller. 
\end {enumerate}

\begin{figure}
\includegraphics[width=78mm]{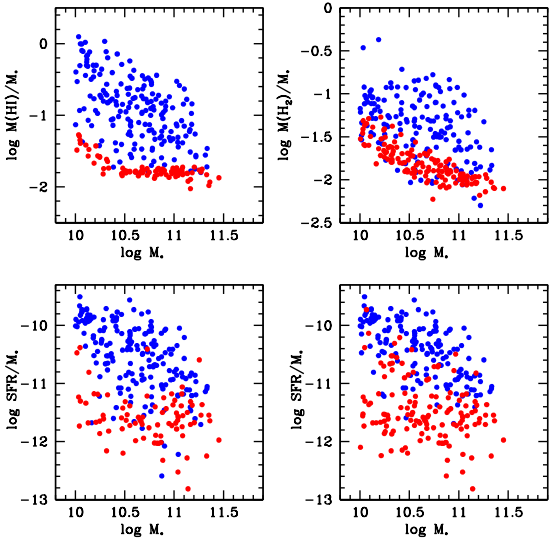}
\caption{ In the left panels, HI mass fraction (log M(HI)/M$_*$)
and specific star formation rate (log SFR/M$_*$) are plotted as  a function
of the $\log M_*$. Blue points denote galaxies with HI line detections.
Red points denote galaxies where HI was not detected.   
In the right  panels, H$_2$ mass fraction (log M(H$_2$)/M$_*$) 
and specific star formation rate (log SFR/M$_*$) are plotted as  a function  
of the log M$_*$. Blue points denote galaxies with CO  line detections.
Red points denote galaxies where CO  was not detected.
\label{sample}}
\end{figure}

In this way, we are able to
classify galaxies in the simulations as HI/CO  ``detections'' or ``non-detections''
and compare properties such as stellar masses, stellar surface densities,
specific star formation rates and concentration indices  with
those of the real galaxies in our survey. We note that stellar masses and star formation rates
are standard outputs of  semi-analytic models. 
Because the F10 models are able to track radial profiles of the gas and the stars
in galaxies, they are also able to  
predict stellar surface mass densities, which are defined as
$0.5 M_*/(\pi R_{50}^2)$, where $R_{50}$ is the radius containing half the
stellar mass of the galaxy.

In semi-analytic models, bulges form either as a result of a merger between
two galaxies, or as a consequence of instabilities that set in when disks
reach a certain critical threshold density.
In the models, the total mass of the bulge is more  
reliably predicted than its size. 
As shown in Figure 1 of Weinmann et al (2009), there is
a tight correlation between the concentration index $C$ (defined as the ratio of
the radii containing 90\% and 50\% of the $r$-band light) and galaxy bulge-to-disk
radio derived using  2-dimensional decomposition codes (Gadotti  2009). 
We have therefore elected to transform the bulge-to-disk ratios predicted by
the models to concentration indices using a fit to this relation: 
C=1.8+2.14$B/T$.
   
\section {Results}

In the analysis presented in this paper, we will attempt to answer two  questions:
\begin {enumerate}
\item  Can the simple disk formation models described in the F10 paper
explain observed gas fraction scaling relations for  galaxies
with gas?  
\item Does the transition between the population of galaxies with gas and
the population without gas occur in the same way in the models and in the data?
\end {enumerate}

We have chosen the gas fraction
detection threshold of our two surveys as the nominal division between the
population of galaxies we will henceforth refer to as ``active'' and 
the population that we will call ``quenched" . Note that
it is possible   that some fraction 
of galaxies classified as ``active'' are actually 
losing their gas and evolving into quenched systems. Conversely, some fraction
of galaxies classified as ``quenched'' may have simply run out of gas 
just prior to the next accretion episode. 
Such galaxies may be thought of as
``transition'' systems. 

In the bottom panels of Figure 1, we plot 
specific star formation rate versus stellar mass for the galaxies in our sample.
In the left panel, galaxies with HI detections are colour-coded blue and
those where the HI line was not detected are colour-coded red. In the
right panel, we do the same thing according to whether the CO line
was detected or not. As can be seen, galaxies with log SFR/M$_* < -11$  
are  not usually detected in HI and almost never in
CO.  This means that there is a rather
clean distinction between ``active'' 
and ``quenched'' galaxies that applies to both
gas fraction and specific star formation rate  
for the majority of galaxies in our sample.
In section 3.2.2, we return to the issue
of whether we can identify a sub-population of galaxies that is in the
{\em process} of being quenched.    
 
\subsection {Gas scaling relations for the active population} 

In Figures 2 and 3 we compare the mean HI and H$_2$ gas fractions
as a function of $M_*$, $\mu_*$, $C$ and SFR/$M_*$ for data and for models.  
The reader is referred to Catinella et al (2010) and Saintonge et al (2011a)
for details on how HI and CO line fluxes are transformed into HI and H$_2$
masses. We note that we have adopted a constant Galactic conversion factor
$\alpha_{\rm CO} = 3.2 M_{\odot}$ (K km s$^{-1}$ pc$^2$)$^{-1}$, which does not
include a correction for the presence of helium. A constant conversion factor should
be a reasonable approximation for the galaxies in our survey, which all
have stellar masses greater  than $10^{10} M_{\odot}$ and gas-phase
metallicities near solar (Saintonge et al 2011b).

In Figures 2 and 3, the blue points show the mean values of M(HI)/M$_*$
and M(H$_2$)/M$_*$ calculated using the galaxies with HI and CO line detections.
Error bars are calculated using bootstrap resampling. The red and black curves indicate
the mean values calculated from the model galaxies that are classified as
detections (see section 2). The black curves show results from the F10 model
that uses the KMT  prescription for  
the conversion of atomic to molecular gas. The red curves show results from the F10 
model that uses the BR pressure-based prescription.
We note that there are many more simulated galaxies than observed galaxies.
In generating the curves, we have ordered the galaxies by 
increasing $M_*$,$\mu_*$, $C$ and SFR/$M_*$ and
computed means for bins of 70 galaxies.      

Figures 2 and 3 show that the models 
reproduce the observed scalings between  mean HI/H$_2$ mass fraction and  
galaxy stellar mass,  surface mass density and concentration quite well. The  differences
between the HI mass fractions  predicted by the KMT and BR prescriptions are small. 
Given the systematic uncertainties in the
calibration of the BR prescription, the fact that the KMT prescription is
based on simple analytic calculations, the fact that our semi-analytic disc formation models
are extremely idealized, and the fact that the model free parameters 
were not tuned to fit the COLD GASS scaling relations,  we find it remarkable how
well these results  agree with the data.
  
The relation between H$_2$ mass fraction and specific star formation
is {\em shallower} than predicted by the models, which  
lie on a relation with slope unity. 
The discrepancy in the relation between M(H$_2$)/M$_*$ and SFR/$M_*$
is easily understood in light of the results presented in Saintonge
et al (2011b). The F10 models  adopt the assumption that $\Sigma_{\rm SFR}
= \alpha \Sigma_{H_2}$ with $\alpha$ constant (corresponding to an
effective molecular gas depletion timescale of 2 Gyr) in all galaxies.
This assumption was motivated by the results in Leroy et al (2008). 
Saintonge et al (2011) showed, however, that there was more than a factor of 5 
variation in the {\em global} molecular gas depletion timescale in 
different galaxies, and that the least actively star-forming galaxies
(i.e lowest values of SFR/$M_*$) had the longest depletion times. This explains   
the shallow observed relation in the bottom right panel of Figure 3.

In Figure 4, we show scatter plots of log SFR/M$_*$ versus  
log M(H$_2$)/M$_*$ and log M(HI)/M$_*$ for our survey galaxies (blue points)
and for galaxies from our mock catalogue (black points).
Results are only shown for the KMT 
atomic-to-molecular gas prescription, because results for the pressure
prescription are virtually the same. As can be seen the models reproduce
the scatter in specific star formation rate versus atomic gas fraction
quite well, but not in specific star formation rate versus molecular
gas fraction. Clearly, there are additional processes at work that
determine the rate at which molecular gas will form stars.
This is the subject of a future paper (Saintonge et al, in preparation).

\begin{figure}
\includegraphics[width=78mm]{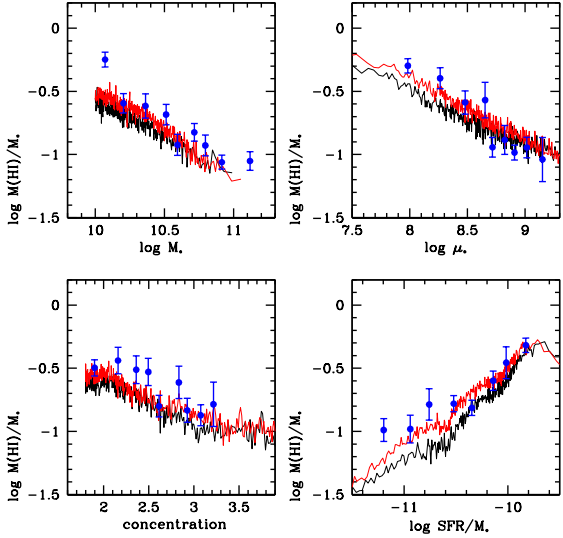}
\caption{ The mean HI mass fraction (log M(HI)/M$_*$) for COLD GASS  galaxies
with HI line detections (blue points) is plotted  as a function of stellar mass,
stellar mass surface density, concentration and specific star formation rate.
Error bars are calculated using bootstrap resampling. Results for
the F10 models are shown as  black (KMT
atomic-to-molecular gas prescription) and red (BR
pressure prescription) curves. 
\label{detecth1}}
\end{figure}

\begin{figure}
\includegraphics[width=78mm]{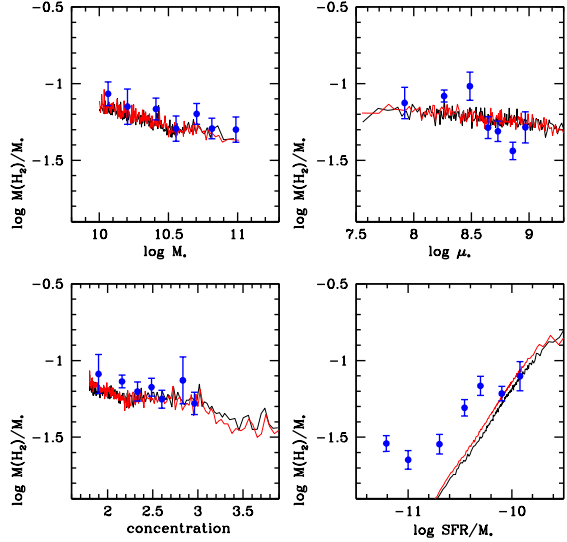}
\caption{ The mean H$_2$ mass fraction (log M(H$_2$)/M$_*$) for COLD GASS  galaxies
with CO line detections (blue points) is plotted  as a function of stellar mass,
stellar mass surface density, concentration and specific star formation rate.
Error bars are calculated using bootstrap resampling. Results for
the F10 models are shown as  black (KMT
atomic-to-molecular gas prescription) and red (BR
pressure prescription) curves. 
\label{detecth2}}
\end{figure}

As well as looking at mean  HI and H$_2$ mass fraction relations for
galaxies with detections, it is instructive to analyze 
{\em full distribution functions} of gas mass fractions. These are illustrated in Figures
5 and 6. The data are shown as blue histograms, with error bars computed from bootstrap resampling.
Black and red curves 
are for the models, as before. We plot distributions of
log M(HI)/M$_*$ and log M(H$_2$)/M$_*$ in three different intervals
of stellar mass,  stellar mass surface density and concentration. 
These plots show that the models
provide a good representation not only of the mean HI/H$_2$ mass fractions, but also of 
the {\em scatter around the mean}. As discussed in F10 and will be illustrated below,
there is a tight relation between stellar mass and dark matter halo mass
in the models.  The scatter around the mean
gas fraction at fixed dark matter halo mass is set by a) the spin parameter of the halo, which
determines the contraction factor of the accreted gas and hence the rate at which
it will be consumed into stars, and b) the recent gas accretion history of the galaxy.
The fact that the scatter in the models and data agree so well 
lends support to this basic picture.

We note that in recent years, there have been a number of papers questioning the treatment of disk formation
in semi-analytic models (e.g. Sales et al 2009). The primary objection is the assumption that
the angular momentum of a galaxy expressed in units of that of its surrounding halo ($j_d= J_{gal}/J_{vir}$) , correlates with
mass of the galaxy divided by the mass of its dark matter halo ($m_d=M_{gal}/M_{vir}$)  in a manner 
that is insensitive to feedback, i.e. $j_d=m_d$ is assumed for all halos.
In our scheme this is {\em not} the case. We only assume that $j_d=m_d$ for the gas that is cooling 
instantaneously at any given time in any given halo.  Because we track the growth of galaxies and  dark matter halos
with time, and because feedback effects will affect the mass of gas that cools in halos of
different masses at different redshifts, the resulting 
relation between the angular momentum of the galaxy and its  mass fraction
will depend  on feedback.

Figure 5 also shows that the models   
miss a minority population of galaxies  with high atomic gas fractions and with
low stellar masses, surface mass densities and concentrations.
What are these missing galaxies? In three recent papers, we have studied the
properties of galaxies in our survey that contain significantly more atomic gas than
would be predicted from their UV/optical colours and sizes 
(Moran et al 2010; Wang et al 2011; Moran et al 2011). 
The main conclusion is that such galaxies have very blue outer disks with
young stellar populations and low gas-phase metallicities.
These results led us to propose that these unusually HI-rich systems have experienced
a recent gas accretion event, resulting in growth of the outer disk.  

In the F10 models,
the accreted gas is always added to the disk with an
exponential surface density profile.
In real galaxies,
the accreted gas  may initially be distributed 
in the outer regions of the galaxy.
Dynamical
perturbations in the form of spiral density waves, interaction with companions or with the
non-axisymmetric gravitational potential of the surrounding dark matter halo,
will eventually the cause the gas to flow inwards and reach high enough densities to
form molecular clouds. 
If we change the gas accretion prescriptions so that accreted gas is initially dumped
in the outskirts of the disk, this may      
produce a tail of HI-rich galaxies. 
Observations of SFR, metallicity and gas {\em profiles}  will be required  to constrain
models of this nature.

\begin{figure}
\includegraphics[width=78mm]{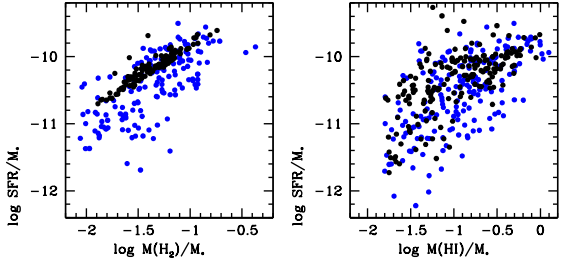}
\caption{ Scatter plots of  H$_2$ mass fraction (log M(H$_2$)/M$_*$,left) 
and HI mass fraction (log M(HI)/M$_*$, right) for COLD GASS  galaxies
are shown as blue points. Results from the F10 models
with KMT atomic-to-molecular gas prescription  are shown as black points. 
\label{sfgas}}
\end{figure}

\begin{figure}
\includegraphics[width=84mm]{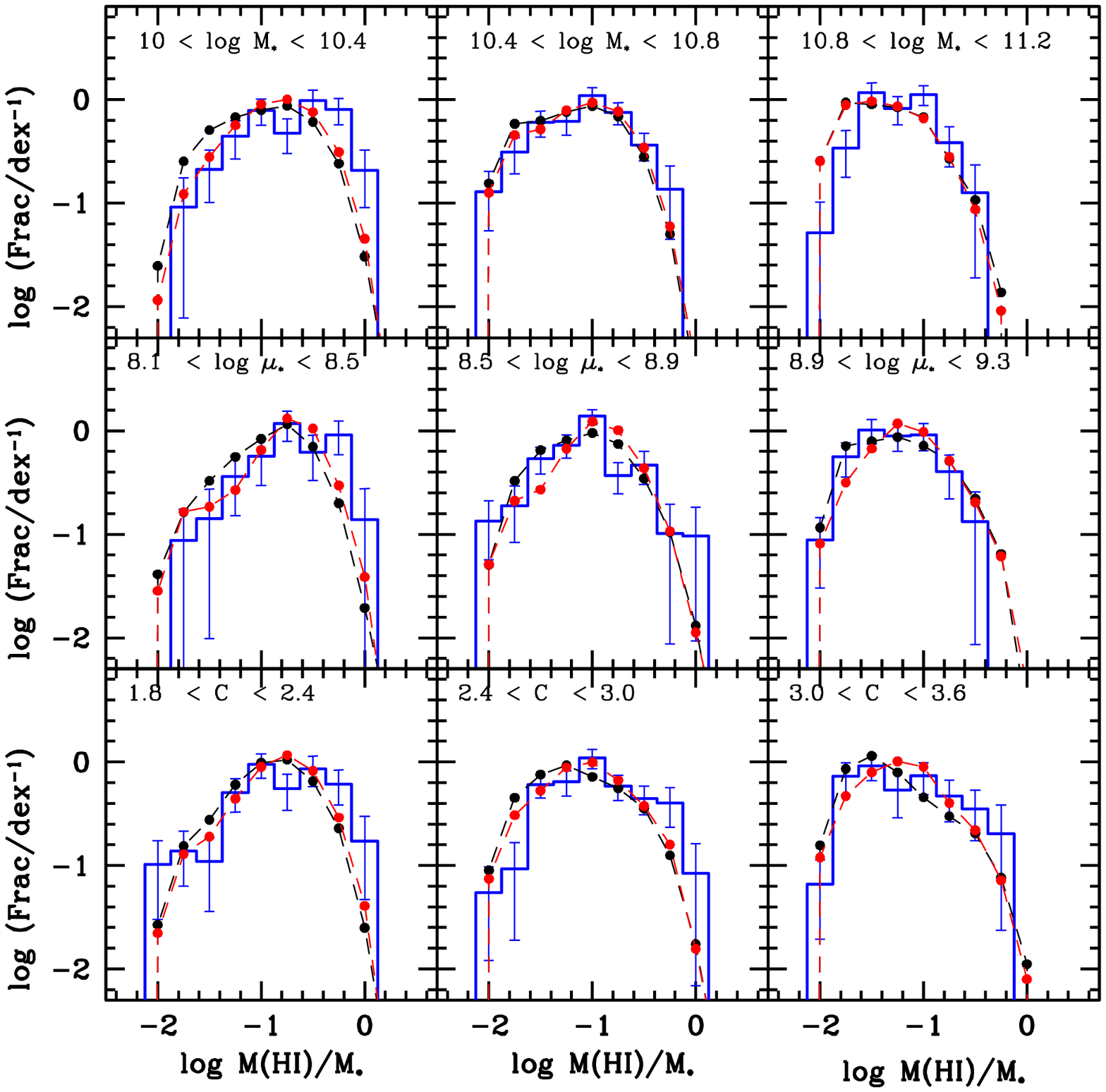}
\caption{ Distribution functions of  HI mass fraction (log M(HI)/M$_*$) for COLD GASS  galaxies
with HI line detections in 3 ranges of  stellar mass,
stellar mass surface density and concentration  (blue
histograms).
Error bars are calculated using bootstrap resampling. Results for
the F10 models are shown as  black and red curves. The colour-coding of the model
curves has the same meaning as in Figures 2 and 3. 
\label{fracfunh1}}
\end{figure}

\begin{figure}
\includegraphics[width=84mm]{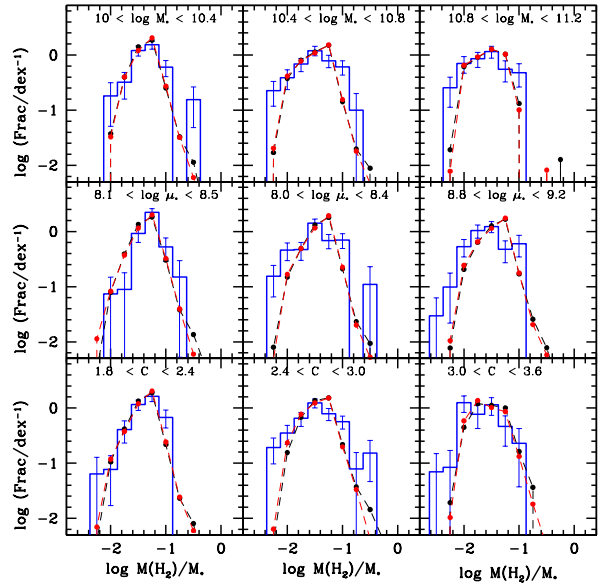}
\caption{ Distribution functions of  H$_2$ mass fraction (log M(H$_2$)/M$_*$) 
for COLD GASS  galaxies
with CO  line detections in 3 ranges of  stellar mass,
stellar mass surface density and  concentration (blue
histograms) .
Error bars are calculated using bootstrap resampling. Results for
the F10 models are shown as  black and red curves. The colour-coding of the model
curves has the same meaning as in Figures 2 and 3. 
\label{fracfunh2}}
\end{figure}

\subsection {The nature of the quenched population}   
In this section, we investigate whether the ``transition'' between the population
of galaxies with detectable atomic and molecular gas and the populations with
gas fractions below $\sim 0.015$ occurs in the same way in models as in the data.  
In Figures 7 and 8, we plot the fraction of HI/CO-detected galaxies as a function
of stellar mass, stellar mass surface density, concentration index and specific
star formation rate. Results from the survey are shown as solid blue lines. The
dashed blue lines indicate the $\pm 1 \sigma$ uncertainty in the detected 
fraction as a function of these parameters. Models results are shown 
in red and black (the colour-coding is the same as  in Figures 2 and 3).   
The main conclusion from these two figures is that the models and the data do
not agree.  First, the dependence of the
detected fraction on stellar mass is
weaker in the data than in the models.
This is true for both the HI and the CO-detected fractions.        
Second, the fraction of CO-detected galaxies drops very strongly as a function of
both stellar surface density and concentration in the real data. 
This is not seen in the models.

\begin{figure}
\includegraphics[width=78mm]{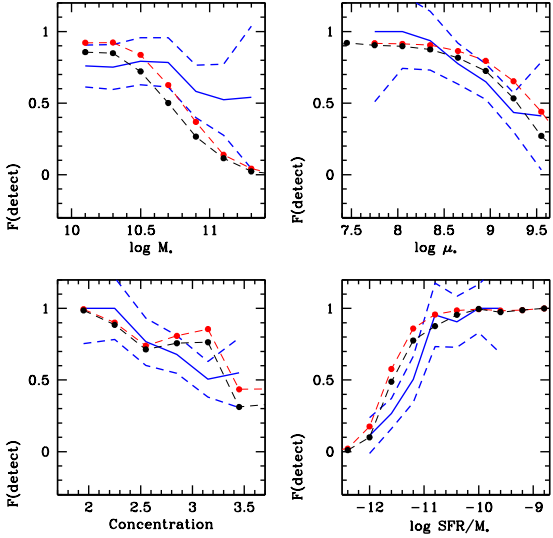}
\caption{ The fraction of COLD GASS  galaxies
with HI line detections  is plotted  as a function of stellar mass,
stellar mass surface density, concentration and specific star formation rate (blue curve).
Poisson errors are shown as blue dashed lines. We note that these are computed
for independent bins, so the 1$\sigma$ errors can be read off directly. Results for
the F10 models are shown as  black (KMT
atomic-to-molecular gas prescription) and red (BR
pressure prescription) curves. 
\label{nodetecth1}}
\end{figure}

\begin{figure}
\includegraphics[width=78mm]{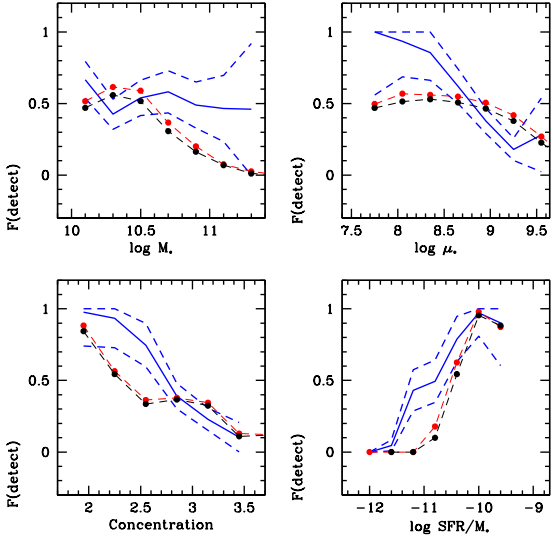}
\caption{ The fraction of COLD GASS  galaxies
with CO line detections  is plotted  as a function of stellar mass,
stellar mass surface density, concentration and specific star formation rate (blue curve).
Poisson errors are shown as blue dashed lines. Results for
the F10 models are shown as  black (KMT
atomic-to-molecular gas prescription) and red (BR
pressure prescription) curves. 
\label{nodetecth1}}
\end{figure}

It is also instructive to look at the distribution of detected and non-detected
galaxies in two-dimensional planes of  stellar mass and galaxy structural
parameters. This is illustrated in Figures 9 and 10.
We plot HI/CO detected galaxies in blue and HI/CO non-detected galaxies in red in 
the  stellar surface density versus stellar mass, 
concentration versus stellar mass, and concentration versus stellar
surface density planes. Results from the survey are shown in the top panels and results
for ``mock catalogues'' of the same size generated from the simulations
are shown in the bottom panels. In the 
data,  there are clear
{\em thresholds} in both $\mu_*$ ($\sim 3 \times 10^8 M_{\odot}$ kpc$^{-2}$) and
$C$ ($\sim 2.6$), that demarcate the location of  almost all galaxies without detectable gas. 
No such threshold is seen in stellar mass $M_*$. This is true for {\em both}
the HI and the CO non-detections.

In the models, the parameters that most clearly demarcates the location of the 
HI non-detections are  stellar mass surface density and stellar mass.  
It is the behaviour of the CO
non-detections, however, that is most discrepant with the models.  
{\em Galaxies without $H_2$ are not  confined to a specific  
location in  structural parameter space
in the same way as in the data.}

We also note that the tight correlation
between stellar surface density and concentration index seen in the top right
panels of both figures, is not present in the models. In addition, the distribution of
concentration indices in the models extends to much lower values than in the real data.
The tight correlation between $C$ and $\log \mu_*$ tells us that in the real Universe, galaxies with larger
bulge-to-disk ratios have disks with higher stellar surface densities.

In the F10 models, the effects of gas inflows on the disk are not taken
into account, which may explain why there is no clear  relation between bulge-to-disk
ratio and stellar surface density.
In a scenario where bulges form when gas flows inwards as a result
of bar-driven inflows or other dynamical instabilities, the gas 
flows will not  only form the bulge, but also  
increase the stellar surface mass density in the inner disk. 

\begin{figure*} \centering
\includegraphics[angle=-90,scale=0.47]{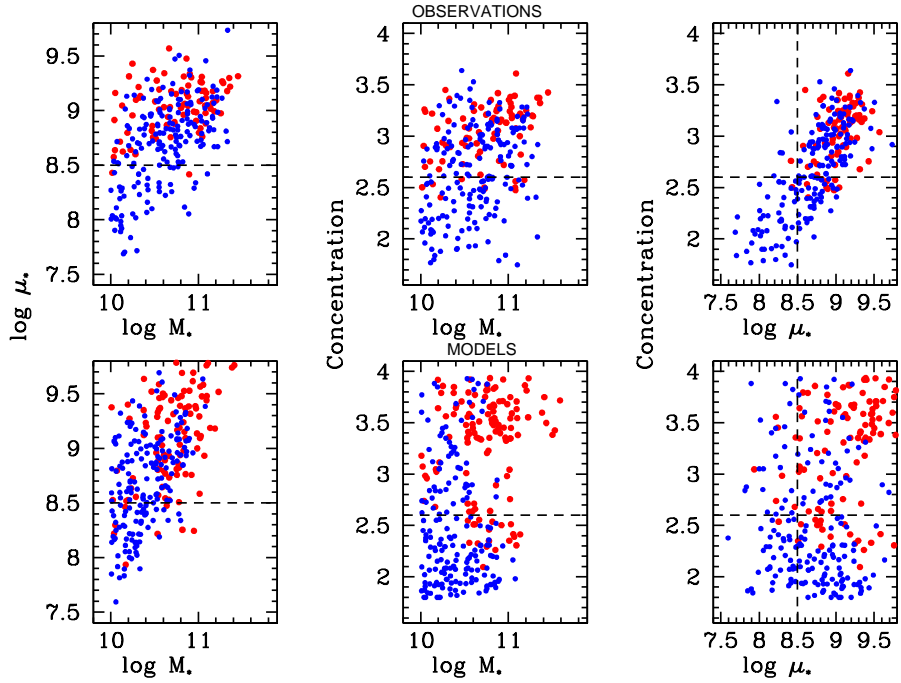}
\caption{ 
Top: COLD GASS  galaxies are plotted in the 2-dimensional planes of 
stellar surface density versus stellar mass, concentration versus stellar
mass and concentration versus stellar surface density. Galaxies with HI line
detections are plotted in blue, while those without HI line detections are
plotted in red. Bottom: Detections and non-detections from
from our F10 model ``mock catalogues'' are plotted
in the same 3 planes.} 
\end{figure*}

\begin{figure*} \centering
\includegraphics[angle=-90,scale=0.47]{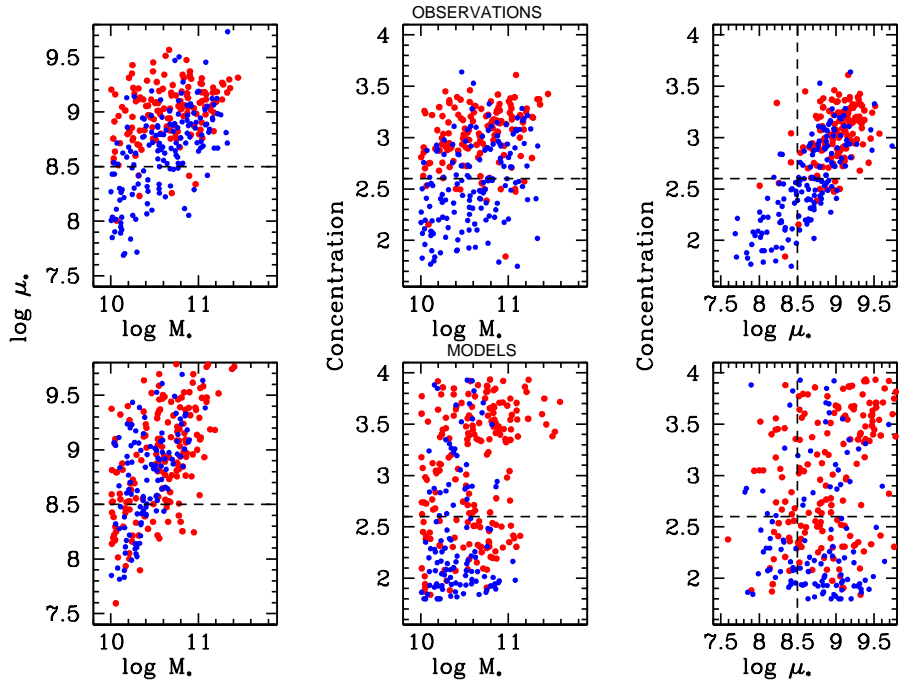}
\caption{ 
Top: COLD GASS  galaxies are plotted in the 2-dimensional planes of 
stellar surface density versus stellar mass, concentration versus stellar
mass and concentration versus stellar surface density. Galaxies with CO line
detections are plotted in blue, while those without CO line detections are
plotted in red. Bottom: Detections and non-detections from
from our F10 model ``mock catalogues'' are plotted
in the same 3 planes.} 
\end{figure*}

\subsubsection{Origin of quenching thresholds in the models}

We now elucidate the origin of the strong trend in the fraction of quenched galaxies as a function of stellar mass  
that is seen for  the model galaxies.
In current semi-analytic models, there are two physical processes that
remove the supply of new gas to a galaxy and shut down star formation.

\begin {enumerate}
\item {\em Radio Mode Feedback.} Black holes are able to grow by accreting
hot gas from the surrounding halo. The growth rate is given by
$ \dot{M}_{\rm BH} = \kappa (f_{hot}/0.1)(V_{vir}/200 {\rm km/s})^3
(M_{\rm BH}/10^8 M_{\odot}) M_{\odot}/{\rm yr}$ (Croton et al 2006; Guo et al 2011),
where $f_{hot}$ is the ratio of hot gas mass to dark matter mass in the surrounding halo
or sub-halo, $V_{vir}$ is the virial velocity of the halo, $M_{BH}$ is the black 
hole mass, and $\kappa$ is an efficiency parameter. Some fraction of the rest
mass energy of the accreted material is assumed to be transferred to the surrounding
hot gas by radio jets. The models assume an energy input rate
$\dot{E}_{\rm radio} = 0.1 \dot{M}_{\rm BH} c^2$, where $c$ is the speed of light.  
This leads to a reduction in the cooling rate of hot gas of the form
$\dot{M}_{\rm cool, eff} = \dot{M}_{\rm cool} - 2 \dot{E}_{\rm radio}/V_{\rm vir}^2$.    
\item {\em Gas stripping.} When a galaxy is accreted by a more massive dark matter
halo, it becomes a ``satellite''. In the models of De Lucia \& Blaizot (2007), the
hot gas surrounding the satellite is stripped instantaneously, leading to a 
sharp reduction in the cooling rate onto the satellite.
In the more recent models of Guo et al (2011), the dark matter and hot gas surrounding the
satellite are removed more gradually  both by tidal forces and by ram-pressure stripping.
\end {enumerate}

The parameter $\kappa$ is tuned to reproduce the high mass end of the stellar
mass function.
The net effect is  illustrated in Figure 11 where we plot 
galaxies from our model catalogue in the 2D  planes of
subhalo mass  versus $ \log M_*$ , subhalo mass 
versus $ \log \mu _*$, and subhalo mass versus $C$. Galaxies that
are predicted to be detected in CO are colour-coded blue, while those
predicted to be non-detections are colour-coded red.   
Figure 11 shows a clear transition between active and quenched galaxies 
at a subhalo mass
of $10^{12} M_{\odot}$, which is  a factor of $\sim 3$ larger than the
mass where dark matter halos are predicted to  transition to hosting  
a static halo of hot gas (Birnboim \& Dekel 2003; Dekel \& Birnboim 2006; Croton et al 2006).  
The reason why the transition is very sharp in halo mass is because    
the primary dependence of the mass
that is accreted by the black hole is on halo
virial velocity and not on black hole mass. 
We note that there is an additional population of quenched galaxies 
in low mass subhalos.
These are the  satellite galaxies, where the subhalo
has been stripped by tidal forces. As can be seen, they form a small
minority of the quenched population in  the stellar mass range of the
galaxies in the GASS survey.

\begin{figure*} \centering
\includegraphics[angle=-90,scale=0.47]{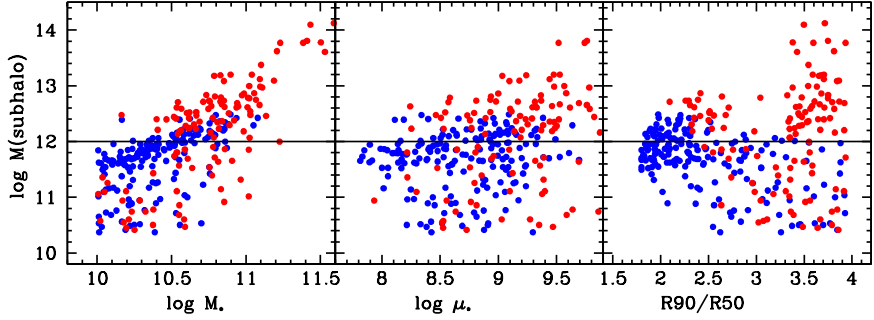}
\caption{ 
F10 model galaxies are plotted in the 2-dimensional planes of 
subhalo mass versus stellar mass, subhalo mass versus stellar surface density  and
subhalo mass versus concentration. Galaxies predicted  to have CO line
detections are plotted in blue, while those predicted not to have CO line detections are
plotted in red.} 
\end{figure*}

Figure 11  clearly illustrates {\em why} quenching  in the model galaxies
is most strongly dependent on stellar mass and surface density, 
and largely independent of 
concentration. This is because in the model, stellar mass
and surface density correlate with the virial mass of the host halo, but 
concentration does not. There is direct observational evidence in
support of these results. Mandelbaum et al (2006) used weak gravitational lensing to
derive halo masses of galaxies as a function of stellar mass and morphology.
They found  stellar mass to be a good proxy for halo mass. For a given halo mass,
the stellar mass was independent of concentration/morphology below $M_* = 10^{11} M_{\odot}$.

\subsubsection{Origin of quenching thresholds in real galaxies}

In the data, quenching thresholds are seen as a 
function of stellar surface density and concentration, but not as a function of
stellar mass. 
Even for galaxies with  stellar masses as high as $10^{11} M_{\odot}$,
which are observed to reside in halos with masses in the range $3 \times 10^{12} -10^{13} M_{\odot}$,
HI and CO are generally still detected if concentrations and densities are low.
This strongly suggests
that  {\em processes associated with bulge formation must be responsible for
shutting off the  gas supply in galaxies.} 

We note that this hypothesis was already
put forward by Kauffmann et al (2006), based on an analysis of the {\em scatter} in 
the colours and spectral properties of galaxies as a function of stellar mass,
stellar surface density and concentration. 
Kauffmann et al (2006) made an ansatz that 
scatter in specific star formation rate
reflected scatter in gas content. The fact that  
specific star formation rates as well as their scatter decreased sharply 
above a characteristic density threshold of $3 \times 10^8 M_{\odot}$ kpc$^{-2}$
and concentration index of $2.6$
and that this threshold was largely independent of the stellar mass of the galaxy,
was taken as evidence that gas accretion was no longer 
occurring in bulge-dominated systems. 
The COLD GASS and  GASS surveys have now demonstrated
that the neutral
gas content of galaxies is weakly dependent on stellar mass and  decreases sharply  
at $\mu_* >  3 \times 10^8 M_{\odot}$ kpc$^{-2}$ and $C>2.6$.

We have not yet ascertained {\em why} bulge-dominated galaxies no longer
accrete gas and form stars efficiently. Figure 3 shows that galaxies with
$\mu_* >  3 \times 10^8 M_{\odot}$ kpc$^{-2}$ have molecular gas mass fractions that are
slightly depressed relative to model predictions. 
Could such galaxies be transitioning to the red sequence
on short timescales as proposed by
Schawinski et al (2009)?

We note that molecular gas is generally concentrated
towards the inner regions of galaxies.      
The SDSS fiber spectra probe the central 1-2 kpc central regions of the
GASS galaxies and contain diagnostics of the recent star formation history
in the bulge.     
In Figure 12, we plot  
H$\delta_A$ as a function of D$_n$(4000) for the galaxies in the survey.
Galaxies where CO was not detected are colour coded red, those with detections
and $\log \mu_* > 8.6$ are colour coded-blue, and those with detections and
$\log \mu_* < 8.6$ are colour-coded cyan. As discussed in detail in
Kauffmann et al (2003a), by combining these two stellar absorption line
indices, we can diagnose if the recent star formation
histories of a population of galaxies have been smooth  or 
``bursty'' on average. Starbursts of duration less than 1-2 hundred million years,
displace  a significant fraction of galaxies to higher values of
H$\delta_A$ at a given value of D$_n$(4000). Likewise, if star formation is truncated
over a timescale of less than a few hundred million years, galaxies will
also be displaced to higher than average values of H$\delta_A$  for
around a Gyr following the truncation event (see for example, Kauffmann et al 2004).

\begin{figure}
\includegraphics[width=78mm]{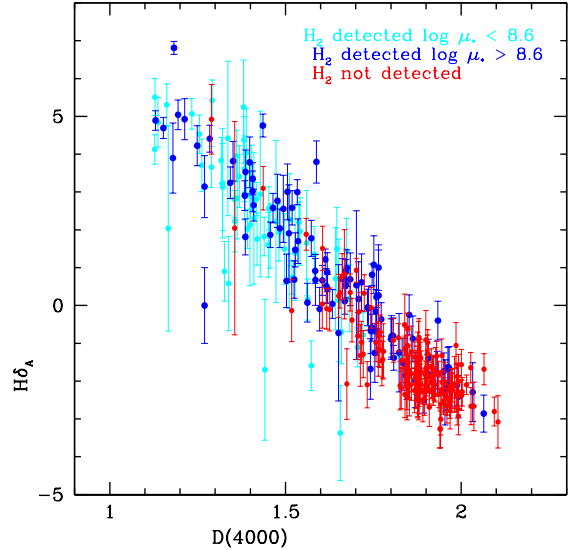}
\caption{ The Lick H$\delta_A$ index is plotted as a function
of 4000 \AA\ break strength for  COLD GASS with
CO line detections and $\log \mu_* > 8.6$ (blue). CO line detections
and $\log \mu_* < 8.6$ (cyan) and for COLD GASS galaxies with no
CO line detections (red). The errors on each measurement are also indicated
for each galaxy.
\label{nodetecth1}}
\end{figure}

As can be seen, 
most  GASS galaxies lie
on the same locus in the plane of H$\delta_A$ versus
D$_n$(4000). There is no evidence of any net displacement as a function of  
stellar surface density or as a function of $H_2$ mass fraction, indicating that
the star formation histories of most  high surface density galaxies with gas 
have been smooth.                
We note, however, that 
 galaxies with CO line detections and
high stellar surface densities {\em do span the largest range in stellar population
parameters}. The have D$_n$(4000) values as low as those of the  
low-density H$_2$-rich   population, and as high as those of the  
``quenched'' population. We also note that the three post-starburst galaxies
that are clearly displaced to higher valued of H$\delta_A$, 
are all detected in $H_2$ have have high stellar surface densities. All of them have unusually high
molecular-to-atomic gas ratios, and one is clearly the end product of a recent
merger (tidal features are visible in the SDSS image). 

This region of parameter space is clearly quite complex, and may consist of 
several distinct sub-populations of galaxies on different evolutionary trajectories.
If so, additional data will be required  to pull these apart. 
In particular, studies 
of the {\em distribution and the kinematics of the gas} 
may shed significant light on its
origin and its eventual fate.

\section {Summary}
     
We compare the semi-analytic models of galaxy formation of Fu et al. (2010),
which track the evolution of atomic and molecular gas in galaxies, with
gas fraction scaling relations derived from a
stellar mass-limited sample of 299 galaxies from the COLD GASS
survey. These galaxies have measurements of the  CO(1-0) line 
from the IRAM 30-m telescope
and the HI line from Arecibo, as well as measurements of stellar masses, structural
parameters and star formation rates derived from GALEX+SDSS photometry.   

Our analysis addresses two key  questions: 1) Can the semi-analytic 
disk formation models
explain the observed gas fraction scaling relations in galaxies
with gas and ongoing star formation?, 2) Does the transition between   
the population of galaxies with gas and
the ``quenched'' population without gas occur in the same way 
in the models and in the data?

In answer to the first question,
we conclude that the disk models provide a reasonable description of
how condensed  baryons are partitioned into stars, atomic gas and molecular
gas as a function of galaxy  mass and size. Trends as a function of bulge-to-disk ratio
are not well reproduced. In particular, the models do not account for the fact
that at fixed stellar mass, there is a  tight relation between
the size of a galaxy and its bulge-to-disk ratio.
 
In answer to the second question, we conclude that our data 
disagree with the current {\em implementation} of radio-mode feedback in the models.
In the models,  the  observable parameter that best predicts whether
a galaxy has been quenched is its stellar mass. Our data shows that
the fraction of quenched galaxies is largely independent of  stellar mass, 
but depends strongly on galaxy
bulge-to-disk ratio and stellar surface density. In other words, even low mass
galaxies with bulges have high probability of being
quenched.  We conclude that processes associated with 
bulge formation are  
thus likely to be responsible both for depleting the neutral gas in galaxies
and  for shutting off the  
gas supply in these systems.     

Solving these problems will require substantive changes to the
way gas transport and bulge and black hole formation is treated in the models, as well 
as to the way feedback from AGN is implemented. We now outline our best guess
as to how this might work in practice.  

In the current models, newly accreted gas is assumed to have an exponential
profile. If the profile is initially shallower than exponential,
more atomic gas would collect in an largely inert reservoir in the outer
regions of the galaxy. This gas would later be driven towards the center
of the galaxy by dynamical perturbations, where it would form
molecular gas and stars in the inner region of the disk. The inflowing gas would
also contribute to the growth of the bulge. A prescription of this kind
may result in a better correlation between stellar surface 
mass density and bulge-to-disk ratio. 

There is now  
evidence from studies of complete samples of nearby galaxies that bars 
are very common in  disk galaxies.  Barazza, Jogee \& Marinova (2008)
find that 70\% of disk-dominated galaxies host bars. Moreover, barred galaxies
(in particular galaxies with strong bars) are associated with
higher central molecular gas fractions and enhanced  rates of
central star formation 
(Sakamoto et al 1999; Ellison et al 2011; Wang et al 2012, in preparation), suggesting 
that bulges are forming in these systems . Finally, barred galaxies do not
exhibit any axcess of nearby companions, suggesting that they are not triggered by
interactions (Li et al. 2009). 
  
All of this suggests that internally-driven gas transport processes are important in   
the formation of  bulges, but resolved maps of the atomic and molecular 
gas distributions in complete samples of
galaxies and more detailed theoretical calculation of
gas inflow rates in galaxies are required in order to quantify  this in more detail.    

In the current models, radio-mode feedback suppresses the cooling of gas in
halos more massive than $10^{12} M_{\odot}$.  
In addition, in massive, high density galaxies, atomic gas is
transformed very efficiently into molecular gas and 
thence into stars. These two mechanisms appear to be
sufficient to place all ``HI-quenched'' galaxies in the high  
$M_*$ amd  $\mu_*$  corner of parameter space, as seen in the bottom  panel 
of Figure 9.  The same mechanisms 
are not, however, sufficient to place the 
``H$_2$-quenched'' galaxies in this same region of parameter
space. Nor can they explain why galaxies deficient in both HI 
and H$_2$ are almost always  found in galaxies with 
substantial bulge components. The simplest inference is that the
neutral gas in galaxies is either consumed or removed when bulges form,
and that subsequent gas accretion into the atomic phase is suppressed in a significant fraction
of bulge-dominated galaxies. 

More detailed studies of the spatially-resolved  kinematics of the gas 
and how gas motions across  galaxies correlate with the
presence of  accreting black holes, jets and the distribution of star formation,
will be needed before we are able to
pinpoint the actual physical mechanisms 
responsible for removing/depleting the gas in these systems
(see for example Hopkins et al 2011). Quenching may occur over short timescales
during particular phases of bulge formation, so it may be necessary to survey large 
samples before a definitive conclusion can be reached. 
In addition, the mechanisms that
inhibit the late accretion of gas in
bulge-dominated systems remain to be clarified.

\section*{Acknowledgments}
This work is based on observations carried out with the IRAM 30 m telescope. IRAM is supported by INSU/CNRS (France), MPG (Germany), and IGN (Spain). We sincerely thank the staff of the telescope for their help in conducting the COLD GASS observations and Qi Guo for helpful discussions.



\begin{thebibliography}{}

\bibitem[\protect\citeauthoryear{Baldry et al.}{2004}]{2004ApJ...600..681B} 
Baldry I.~K., Glazebrook K., Brinkmann J., Ivezi{\'c} {\v Z}., Lupton 
R.~H., Nichol R.~C., Szalay A.~S., 2004, ApJ, 600, 681 

\bibitem[\protect\citeauthoryear{Barazza, Jogee, 
\& Marinova}{2008}]{2008ApJ...675.1194B} Barazza F.~D., Jogee S., Marinova I., 2008, ApJ, 675, 1194 

\bibitem[\protect\citeauthoryear{Bigiel et al.}{2008}]{2008AJ....136.2846B} 
Bigiel F., Leroy A., Walter F., Brinks E., de Blok W.~J.~G., Madore B., 
Thornley M.~D., 2008, AJ, 136, 2846 

\bibitem[\protect\citeauthoryear{Blitz \& 
Rosolowsky}{2006}]{2006ApJ...650..933B} Blitz L., Rosolowsky E., 2006, ApJ, 650, 933 (BR)  

\bibitem[\protect\citeauthoryear{Bower et al.}{2006}]{2006MNRAS.370..645B} 
Bower R.~G., Benson A.~J., Malbon R., Helly J.~C., Frenk C.~S., Baugh 
C.~M., Cole S., Lacey C.~G., 2006, MNRAS, 370, 645 

\bibitem[\protect\citeauthoryear{Catinella et 
al.}{2010}]{2010MNRAS.403..683C} Catinella B., et al., 2010, MNRAS, 403, 
683 

\bibitem[\protect\citeauthoryear{Cattaneo et 
al.}{2006}]{2006MNRAS.370.1651C} Cattaneo A., Dekel A., Devriendt J., 
Guiderdoni B., Blaizot J., 2006, MNRAS, 370, 1651 


\bibitem[\protect\citeauthoryear{Croton et al.}{2006}]{2006MNRAS.365...11C} 
Croton D.~J., et al., 2006, MNRAS, 365, 11 


\bibitem[\protect\citeauthoryear{De Lucia, Kauffmann, 
\& White}{2004}]{2004MNRAS.349.1101D} De Lucia G., Kauffmann G., White S.~D.~M., 2004, MNRAS, 349, 1101 

\bibitem[\protect\citeauthoryear{De Lucia 
\& Blaizot}{2007}]{2007MNRAS.375....2D} De Lucia G., Blaizot J., 2007, MNRAS, 375, 2 


\bibitem[\protect\citeauthoryear{Dutton, van den Bosch, 
\& Dekel}{2010}]{2010MNRAS.405.1690D} Dutton A.~A., van den Bosch F.~C., Dekel A., 2010, MNRAS, 405, 1690 

\bibitem[\protect\citeauthoryear{Ellison et 
al.}{2011}]{2011MNRAS.416.2182E} Ellison S.~L., Nair P., Patton D.~R., 
Scudder J.~M., Mendel J.~T., Simard L., 2011, MNRAS, 416, 2182 

\bibitem[\protect\citeauthoryear{Fu et al.}{2010}]{2010MNRAS.409..515F} Fu 
J., Guo Q., Kauffmann G., Krumholz M.~R., 2010, MNRAS, 409, 515 (F10)  

\bibitem[\protect\citeauthoryear{Gnedin, Tassis, 
\& Kravtsov}{2009}]{2009ApJ...697...55G} Gnedin N.~Y., Tassis K., Kravtsov A.~V., 2009, ApJ, 697, 55 

\bibitem[\protect\citeauthoryear{Governato et 
al.}{2004}]{2004ApJ...607..688G} Governato F., et al., 2004, ApJ, 607, 688 

\bibitem[\protect\citeauthoryear{Governato et 
al.}{2007}]{2007MNRAS.374.1479G} Governato F., Willman B., Mayer L., Brooks 
A., Stinson G., Valenzuela O., Wadsley J., Quinn T., 2007, MNRAS, 374, 1479 

\bibitem[\protect\citeauthoryear{Guo et al.}{2011}]{2011MNRAS.413..101G} 
Guo Q., et al., 2011, MNRAS, 413, 101 

\bibitem[\protect\citeauthoryear{Hopkins, Quataert, 
\& Murray}{2011}]{2011arXiv1110.4638H} Hopkins P.~F., Quataert E., Murray N., 2011, arXiv, arXiv:1110.4638 

\bibitem[\protect\citeauthoryear{Kauffmann et 
al.}{2003a}]{2003MNRAS.341...33K} Kauffmann G., et al., 2003, MNRAS, 341, 33 


\bibitem[\protect\citeauthoryear{Kauffmann et 
al.}{2003b}]{2003MNRAS.341...54K} Kauffmann G., et al., 2003, MNRAS, 341, 54 

\bibitem[\protect\citeauthoryear{Kauffmann et 
al.}{2004}]{2004MNRAS.353..713K} Kauffmann G., White S.~D.~M., Heckman 
T.~M., M{\'e}nard B., Brinchmann J., Charlot S., Tremonti C., Brinkmann J., 
2004, MNRAS, 353, 713 

\bibitem[\protect\citeauthoryear{Kauffmann et 
al.}{2006}]{2006MNRAS.367.1394K} Kauffmann G., Heckman T.~M., De Lucia G., 
Brinchmann J., Charlot S., Tremonti C., White S.~D.~M., Brinkmann J., 2006, 
MNRAS, 367, 1394 

\bibitem[\protect\citeauthoryear{Keres, Yun, 
\& Young}{2003}]{2003ApJ...582..659K} Keres D., Yun M.~S., Young J.~S., 2003, ApJ, 582, 659 

\bibitem[\protect\citeauthoryear{Krumholz, McKee, 
\& Tumlinson}{2009}]{2009ApJ...693..216K} Krumholz M.~R., McKee C.~F., Tumlinson J., 2009, ApJ, 693, 216 (KMT)  

\bibitem[\protect\citeauthoryear{Lagos et al.}{2011}]{2011MNRAS.416.1566L} 
Lagos C.~D.~P., Lacey C.~G., Baugh C.~M., Bower R.~G., Benson A.~J., 2011, 
MNRAS, 416, 1566 

\bibitem[\protect\citeauthoryear{Leroy et al.}{2008}]{2008AJ....136.2782L} 
Leroy A.~K., Walter F., Brinks E., Bigiel F., de Blok W.~J.~G., Madore B., 
Thornley M.~D., 2008, AJ, 136, 2782 

\bibitem[\protect\citeauthoryear{Li et al.}{2009}]{2009MNRAS.397..726L} Li 
C., Gadotti D.~A., Mao S., Kauffmann G., 2009, MNRAS, 397, 726 

\bibitem[\protect\citeauthoryear{Lu et al.}{2011}]{2011MNRAS.416.1949L} Lu 
Y., Mo H.~J., Weinberg M.~D., Katz N., 2011, MNRAS, 416, 1949 

\bibitem[\protect\citeauthoryear{Malbon et al.}{2007}]{2007MNRAS.382.1394M} 
Malbon R.~K., Baugh C.~M., Frenk C.~S., Lacey C.~G., 2007, MNRAS, 382, 1394 

\bibitem[\protect\citeauthoryear{McNamara 
\& Nulsen}{2007}]{2007ARA&A..45..117M} McNamara B.~R., Nulsen P.~E.~J., 2007, ARA\&A, 45, 117 

\bibitem[\protect\citeauthoryear{Mo, Mao, 
\& White}{1998}]{1998MNRAS.295..319M} Mo H.~J., Mao S., White S.~D.~M., 1998, MNRAS, 295, 319 

\bibitem[\protect\citeauthoryear{Moran et al.}{2010}]{2010ApJ...720.1126M} 
Moran S.~M., et al., 2010, ApJ, 720, 1126 

\bibitem[\protect\citeauthoryear{Moran et al.}{2011}]{2011arXiv1112.1084M} 
Moran S.~M., et al., 2011, arXiv, arXiv:1112.1084 

\bibitem[\protect\citeauthoryear{Navarro 
\& Steinmetz}{1997}]{1997ApJ...478...13N} Navarro J.~F., Steinmetz M., 1997, ApJ, 478, 13 

\bibitem[\protect\citeauthoryear{Obreschkow et 
al.}{2009}]{2009ApJ...698.1467O} Obreschkow D., Croton D., De Lucia G., 
Khochfar S., Rawlings S., 2009, ApJ, 698, 1467 


\bibitem[\protect\citeauthoryear{Oser et al.}{2010}]{2010ApJ...725.2312O} 
Oser L., Ostriker J.~P., Naab T., Johansson P.~H., Burkert A., 2010, ApJ, 
725, 2312 

\bibitem[\protect\citeauthoryear{Robertson 
\& Kravtsov}{2008}]{2008ApJ...680.1083R} Robertson B.~E., Kravtsov A.~V., 2008, ApJ, 680, 1083 

\bibitem[\protect\citeauthoryear{Saintonge et 
al.}{2011}]{2011MNRAS.415...61S} Saintonge A., et al., 2011a, MNRAS, 415, 61 

\bibitem[\protect\citeauthoryear{Saintonge et 
al.}{2011}]{2011MNRAS.415...32S} Saintonge A., et al., 2011b, MNRAS, 415, 32 

\bibitem[\protect\citeauthoryear{Sakamoto et 
al.}{1999}]{1999ApJ...525..691S} Sakamoto K., Okumura S.~K., Ishizuki S., 
Scoville N.~Z., 1999, ApJ, 525, 691 

\bibitem[\protect\citeauthoryear{Sales et al.}{2009}]{2009MNRAS.399L..64S} 
Sales L.~V., Navarro J.~F., Schaye J., Dalla Vecchia C., Springel V., Haas 
M.~R., Helmi A., 2009, MNRAS, 399, L64 

\bibitem[\protect\citeauthoryear{Schawinski et 
al.}{2009}]{2009ApJ...690.1672S} Schawinski K., et al., 2009, ApJ, 690, 
1672 

\bibitem[\protect\citeauthoryear{Schiminovich et 
al.}{2010}]{2010MNRAS.408..919S} Schiminovich D., et al., 2010, MNRAS, 408, 
919 

\bibitem[\protect\citeauthoryear{Somerville et 
al.}{2008}]{2008MNRAS.391..481S} Somerville R.~S., Hopkins P.~F., Cox 
T.~J., Robertson B.~E., Hernquist L., 2008, MNRAS, 391, 481 


\bibitem[\protect\citeauthoryear{Strateva et 
al.}{2001}]{2001AJ....122.1861S} Strateva I., et al., 2001, AJ, 122, 1861 

\bibitem[\protect\citeauthoryear{Wang et al.}{2011}]{2011MNRAS.412.1081W} 
Wang J., et al., 2011, MNRAS, 412, 1081 

\bibitem[\protect\citeauthoryear{Weinmann et 
al.}{2009}]{2009MNRAS.394.1213W} Weinmann S.~M., Kauffmann G., van den 
Bosch F.~C., Pasquali A., McIntosh D.~H., Mo H., Yang X., Guo Y., 2009, 
MNRAS, 394, 1213 

\bibitem[\protect\citeauthoryear{Williams et 
al.}{2010}]{2010ApJ...713..738W} Williams R.~J., Quadri R.~F., Franx M., 
van Dokkum P., Toft S., Kriek M., Labb{\'e} I., 2010, ApJ, 713, 738 

\bibitem[\protect\citeauthoryear{Wuyts et al.}{2011}]{2011arXiv1107.0317W} 
Wuyts S., et al., 2011, arXiv, arXiv:1107.0317 

\bibitem[\protect\citeauthoryear{Young et al.}{1995}]{1995ApJS...98..219Y} 
Young J.~S., et al., 1995, ApJS, 98, 219 



\end{thebibliography}
\end{document}